gsave
newpath
  20 20 moveto
  20 220 lineto
  220 220 lineto
  220 20 lineto
closepath
2 setlinewidth
gsave
  .4 setgray fill
grestore
stroke
grestore
\end{filecontents*}
\RequirePackage{fix-cm}
\documentclass[twocolumn,epjc3]{svjour3}
\smartqed  % flush right qed marks, e.g. at end of proof
\RequirePackage{graphicx}
\RequirePackage{latexsym}
\RequirePackage[numbers,sort&compress]{natbib}
\RequirePackage{url}
\journalname{}
\begin{document}

\title{Cosmological-model-independent tests of cosmic distance duality relation with Type Ia supernovae and radio quasars} %\thanksref{t1}}
\subtitle{}


\author{Yuan He\thanksref{addr1}
        \and
        Yu Pan\thanksref{e1,addr1}
        \and
        Dong-Ping Shi\thanksref{addr2}
        \and
        Shuo Cao\thanksref{e2,addr3}
        \and
        Wen-Jie Yu\thanksref{addr1}
        \and
        Jing-Wang Diao\thanksref{addr1}
        \and
        Wei-Liang Qian\thanksref{addr1,addr4,addr5}%etc.
}

\thankstext{e1}{e-mail: panyu@cqupt.edu.cn}
\thankstext{e2}{e-mail: caoshuo@bnu.edu.cn}


\institute{College of Science, Chongqing University of Posts and Telecommunications,
   Chongqing 400065, China \label{addr1}
           \and
            School of Electronic and Electrical Engineering, Chongqing University of Arts and Science,
             Chongqing \label{addr2}
           \and
             Department of Astronomy, Beijing Normal University,
             Beijing, 100875, China \label{addr3}
            \and
             Escola de Engenharia de Lorena, Universidade de S\~ao Paulo,
             12602-810, Lorena, SP, Brazil \label{addr4}
            \and
             Center for Gravitation and Cosmology, College of Physical Science and Technology, Yangzhou University,
             Yangzhou 225009, China \label{addr5}
}

\date{Received: date / Accepted: date}

\maketitle

\begin{abstract}
In this paper, we investigate the possible deviations of the cosmic distance duality relation (CDDR) using the combination of the largest SNe Ia (Pantheon) and compact radio quasar (QSO) samples through two model-independent approaches. The deviation of CDDR is written as $D_L(z)/D_A(z)(1+z)^{-2}=\eta(z)$ and $\eta(z)=e^{\tau(z)/2}$, with the parameterizations of $F_1$ ($\tau(z) = 2\epsilon_1 z$) and $F_2$ ($\tau(z) = (1+z)^{2\epsilon_2}-1$). Furthermore, in order to compare the two resulting distances, two cosmological-model-independent methods, i.e., the nearby SNe Ia method and the GP method are employed to match the two distinct data at the same redshift. Our
findings indicate that, compared with the results obtained in the literature, there is an improvement in precision when the latest SNe Ia and QSO samples are used. Specially, in the framework of nearby SNe Ia method, the CDDR would be constrained at the precision of $\Delta\epsilon_{1} = 0.013$ in Model $F_1$ and $\Delta\epsilon_{2}=0.018$ in Model $F_2$. Regarding the GP method, one observes that a larger data size would produce more stringent constraints on the CDDR parameters. Therefore, accompanied by further developments in cosmological observations and the analysis methods, our analysis provides an insight into the evidence for unaccounted opacity sources at an earlier stage of the universe, or at the very least the new physics involved.

\end{abstract}

\section{Introduction}           %% first-level sections will be auto-capitalized
\label{sect:intro}

Cosmic distance duality relation (CDDR) is one of the basic relationships in cosmology, in which the luminosity distance and angular diameter distance at the same redshift can be expressed. If three conditions are satisfied: 1. A metric theory of gravity describes spacetime. 2. Photons travel along unique null geodesics. 3. The number of photons is conserved. The luminosity distance $D_L$ and angular diameter $D_A$ at the same redshift can be simply expressed as \citep{Etherington1933}
\begin{equation}
\frac{D_L(z)}{D_A(z)}(1+z)^{-2}=1 .
\label{eq1}
\end{equation}
It is interesting to note that the validity of CDDR is not only a test of the theory of gravity, but also helps us to understand the temperature profile \cite{cao2011the} and the gas mass density profile \cite{cao2016testing} of galaxy clusters.

The test of the CDDR is usually carried out in the form of parameterization $D_L(z)/D_A(z)(1+z)^{-2}=\eta(z)$. $\eta(z)$ represents how the our universe satisfies CDDR, which can be constrained by observational data. The function $\eta(z)$ parameterized in two distinct forms, $\eta(z)=1+\eta_{0}z$ and $\eta(z)=1+\eta_{0}z/(1+z)$ \cite{Holanda2011}. Li et al. \cite{cao2011testing,Li2011} used SNe Ia and two galaxy cluster samples to test the above parameterized $\eta(z)$. Yang et al. \cite{Yang2019} tested $\eta(z)$ based on the simulated data of Einstein Telescope. Fu et al. \cite{Fu2019} tested the CDDR by combining gravitational wave data with baryon acoustic oscillation data. Their results all show that there is no deviation from CDDR at present. In addition, photon number conservation also affects the effectiveness of the CDDR. Considering that the received photon flux decreases in the form of the function, different parameterized forms can be given. e.g., the parameter of the CDDR is constrained by the parametrization function $\tau(z)$ rather than $\eta(z)$, and their relation is $e^{\tau(z)/2}=\eta(z)$~\citep{Lima2011}. Many scholars combine gravitational wave with electromagnetic wave data to constrain $\tau(z)$, because gravitational wave does not need to satisfy the condition of photon number conservation in the process of gravitational wave propagation \cite{Fu2020,Fu2021,Zhou2019}. In this paper, we only use the data of electromagnetic waves (Supernovae and quasars) and the parameterization of optical depth ($F_1$: $\tau(z) = 2\epsilon_1 z$, $F_2$: $\tau(z) = (1+z)^{2\epsilon_2}-1$) to study the effectiveness of CDDR. In this process, we use two methods (nearby and Gaussian process) to eliminate the redshift gap between the two observed data, and use Markov chain Monte Carlo method to constrain the parameters.

So far, research reported does not show that the CDDR relation deviates significantly from the real universe \cite{Liao2016,Lv2016,Fu2017,Ruan2017,Li2018,Lin2018,Ruan2018,Qi2019b}.
In these works, some tests of the CDDR were carried out by adopting specific cosmological models, and others were given in a model-independent way.
When uses the cosmological-model-independent methods, they usually use astronomical observations to perform a test of the CDDR.
Paper~\citep{Liao2013} used one of the model-independent ways named nearby SNe Ia methods.
In this way, one proposed to compare the luminosity distances deprived of the standard candle SNe Ia with the one derived from other observational data at the same redshifts.
It assumed that the CDDR is reflected in the reduction of the photon number so that the type Ia supernovae are the ideal tool to estimate the $D_L$.
After the progress of the measurements of the Hubble parameter $H(z)$, many research combining the $H(z)$ with a different subsample of SNe Ia observations~\citep{Holanda2013,Liao2013}. However, there are some significant errors in this way that can't be ignored. Because $H(z)$ describes the expansion rate of the universe rather than the distance, an important uncertainties will be gained in angular diameter distance by integrating these scattered points \citep{Liao2015}. It shows the importance of taking the correlation between different redshifts into account. More importantly, the error due to the miss-match between the $H(z)$ and SNe Ia sample at the same redshift should be considered. In addition, reference \citep{Liao2013} propose two other model-independent methods to investigate the CDDR. One of them is a non-parametric method that reconstructs the observational data using a Gaussian kernel called the GP method. It supplements the data over redshift to provide more available information for us to test the CDDR. In reference \citep{Liao2013} they smooth SNe Ia data to reconstruct the luminosity distance. Other method is the interpolation method which uses the nearby SNe Ia points to obtain the luminosity distance of SNe Ia points with the $H(z)$ data at the same redshift \citep{Liao2013}. Using the way one can avoid any bias brought by redshift incoincidence between SNe Ia data and $H(z)$.

Many different sources have been used to test the CDDR, and each of them has its advantages or disadvantages. It is rewarding to turn to objects covering a wide redshift range with both angular diameter and luminosity distances measurement. Many efforts have been made to study the possibility of using quasars with multiple measurements for such purpose \citep{Zheng2020,geng2020gravitational-wave,2021EPJC...81..903L}. For instance, Reference \citep{Zheng2020} has applied it to derive luminosity distances to quasars from the non-linear relation between the UV and X-ray emission \citep{Risaliti2019} and angular diameter distances from the angular size-redshift relation of compact radio quasars \citep{Cao2017}. In this paper, we give a try to test the CDDR by combining supernovae Ia samples with compact radio quasars. The reduction of the observed flux from SNe Ia can be represented by a factor $e^{-\tau(z)}$, where $\tau(z)$ is the optical depth related to the absorption of the universe. The distance deprived of SNe Ia is related to QSO's as $D_{L,SN}=D_{L,QSO}e^{\tau(z)/2}$. To compare two luminosity distances need to match two source samples at the same redshift. We use the nearby SNe Ia method to obtain four group samples through redshift matching criterion $\Delta z \leq 0.002-0.005$. Specifically, to have a one-to-one matching between them, we also use the GP method to reconstruct our quasars data.

In section 2, we introduce the observational data. In section 3, we mainly introduce the methodology of attaining two different cosmological distances from quasars measurements and the latest supernovae Ia observations. Then we investigate the constraint on two different parameterizations of the CDDR and present our results. In section 4, we express conclusions.

\section{Observational Data}

\subsection{Compact radio quasar sample}
We use a sample of 120 intermediate-luminosity quasars identified by Cao et al. \citep{Cao2017,Cao2018b}. For its negligible dependence on both source luminosity and redshift ($|\rho|\approx 10^{-4},|n|\approx 10^{-3}$), the compact structure sizes of these quasars with multi-frequency VLBI observations are potentially promising standard rulers \citep{Cao2018a}. The angular size-redshift ($\theta-z$)relation of the compact radio sources at different redshifts $\theta(z)$ which has been widely used in cosmic studies can be written as
\begin{equation}
\theta(z)= \frac{l_m}{D_A(z)}
\label{eq2}
\end{equation}
where $lm$, $D_A(z)$ is the intrinsic metric length of the source, and the angular diameter distance respectively. To use Eq. (\ref{eq2}) for cosmological inference, one needs to calibrate $l_m$. Cao et al. \citep{Cao2017} used a cosmological-model-independent method to derive the linear size of the compact structure $l_m=(11.03\pm0.25)pc$. Kellermann \cite{Kellermann1993} propose the angular size-redshift relation firstly. He tended to estimate the slowdown parameter based on 79 compact sources were gained through using Very Long Baseline Interferometry (VLBI)at 5 GHz frequency \citep{Zheng2020}. And then, Gurvits \cite{Gurvits1994} enlarged the method to estimate the deceleration parameter using 337 active galactic nuclei observed at 2.29GHz by Morabitto et al. \cite{MORABITO1985}. The difference is that Gurvits express the source compactness using the modulus of visibility $\Gamma = \frac{S_c}{S_t}$ and calculate the characteristic angular size of radio sources through function$\theta(z)=2\sqrt{-ln\Gamma ln2}/\pi B_\theta$. Furthermore, Gurvits \cite{Gurvits1994} discussed the influence of dispersion in the relation between $\theta$ and $z$ and used another sample collected from the literature. And then, Naess et al. \cite{Naess2014}who wanted to comprehend the physical meaning of the standard rulers by trying to construct a reasonable model based on the former achievement. The intrinsic length $l_m$ phenomenologically depends on the source luminosity and the redshift as
\begin{equation}
l_m=lL^{\rho}(1+z)^{n}
\end{equation}
where $l$ is the scaling factor of linear size, power-law exponents $\rho$ and $n$ express how deep the intrinsic length depends on source luminosity and redshift, respectively. In 1985, Morabitto et al. \citep{MORABITO1985} detected 917 radio sources in several VLBI studies at 2.29 GHz. Jackson and Jannetta \citep{JACKSON2006} applied the information gained from the NASA/IPAC Extragalactic Database and contemporaneous radio measurements to derive a catalog with 613 objects covering the redshift range from 0.0035 to 3.787. Considering the difference of quasars, BL Lacertae objects and radio galaxies \citep{2015ApJ...806...66C}, Cao et al. identified the sample of 120 quasars by applying the selection criteria of a flat spectral index (a flat segment of the $\theta$-$\alpha$ diagram $-0.38 \leq \alpha \leq 0.18$) and luminosity in the range of $10^{27}W/Hz \leq L \leq 10^{28}W/Hz$ and export result with the angular size in $0.424 \leq \theta \leq 2.734$  milliarcseconds to former sample \citep{Cao2018b}. The external absorbing does not play any significant role in the observed angular sizes at least up to 43 GHz. For the majority of radio sources in our sample, their angular sizes are inversely proportional to the observing frequency \citep{Cao2018a}. Therefore, those compact quasars can be used as standard rulers for cosmological inference with the advantage of having a larger redshift. This sample has been applied in many cosmological analyses, such as the observational constraints on the interaction between cosmic dark sectors \cite{zheng2017ultra}, General Relatively and modified gravity theories \cite{Qi2017,Xu2017,2020ApJ...888L..25C}, the Hubble constant and cosmic curvature \cite{cao2019milliarcsecond,2021MNRAS.503.2179Q}. In this paper, we will use it to generate the angular-diameter distances matched with SNe Ia.

\subsection{Type Ia supernova sample}

In this paper, we consider the newest SNe Ia sample Pantheon, including 1048 SNe Ia in the redshift range $0.01 < z < 2.3$ released by Pan-STARRS1 (PS1) Medium Deep Survey \citep{Scolnic2017}. SNe Ia is the most direct indication of the accelerated expansive universe as 'standards candles'. If the universe is not transparent, the modulus deprived of SNe Ia will be influenced systematically and its luminosity distance will be increased. If $\tau(z)$ describes the decrease caused by any factor reducing the photon number,the received flux could be expressed as $(e^{-\tau(z)})$. According to CDDR, we can replace the luminosity distance given by supernovae $(D_{L,SN})$ with the luminosity distance given by quasars $(D_{L,QSO})$:
\begin{equation}
D_{L,SN}=D_{L,QSO}e^{\tau(z)/2}
\label{eq3}
\end{equation}

Note that the distance modulus could provide the opacity-dependent luminosity distance as $\mu(z)=5 log_{10}(D_{L}) + 25$. Therefore the theoretical distance modulus could be obtained as
\begin{equation}
\mu_{th}(z)=5log_{10}(D_{L,QSO})+25+2.5log_{10}(e)\tau(z)
\end{equation}

\section{Constraints on the cosmic opacity parameters through two cosmological-model-independent methods}
\label{sect:methods}
In the process of light propagation, the photon number may not be conserved due to the scattering and absorption of some opaque sources. Therefore, considering an opaque universe, the photon flux received by the observer will be reduced by $e^{-\tau(z)}$. If $\tau(z) \neq 0$, it indicates the deviation of CDDR. The main difference between the CDDR and the cosmic opacity is that they choose different parameterized forms for $\eta(z)$. Holanda et al. \cite{Holanda2011} proposed two parameterized forms of $\eta(z)$, $\eta(z)=1+\eta_{0}z$ and $\eta(z)=1+\eta_{0}z/(1+z)$ to test the CDDR. They consider these expressions have several advantages such as a manageable one-dimensional phase space and a good sensitivity to observational data. For cosmic opacity, it is studied from the parameterization of optical depth $\tau(z)$, $\eta(z)=e^{\tau(z)/2}$ as described earlier. One of the effective methods to test the CDDR is to verify the conservation of photon numbers. Therefore, in this paper, we use optical depth parameterization considering photon number conservation to verify the effectiveness of the CDDR.

There are many parameterized functions of $\tau$ that have been presented, and we choose two general functions:
\begin{equation}
F_1: \tau(z) = 2\epsilon_{1}z
\end{equation}
\begin{equation}
F_2: \tau(z) = (1+z)^{2\epsilon_2}-1
\end{equation}

These are two typical parameterizations, and have been widely used \cite{Fu2020,Fu2021}. If $\epsilon_{1}/\epsilon_{2} = 0$, it means that photon number is conserved and the CDDR is valid. Besides, in the subsequent analysis the CDDR will also be studied in individual redshift bins, in order to perform the CDDR test in a more generalized way.
The two functions could be similar essentially when $z\ll1$ and they are both wavelengths-independent on the optical band. But if $z$ is not very small, the second function will differ from the former. The origin of the former function is the parametrization function of CDDR $\eta(z)=(1+z)^{\epsilon}$ with small $\epsilon$ and $z$. Qi et al. \citep{Qi2019} If the likelihood of $\epsilon$ is constrained at the peak of $\epsilon = 0$, the photon is scattered or absorbed rarely, which means the transparent universe.
In this paper, we use two different model-independent methods to constrain the parameter of cosmic opacity. One is the nearby SNe Ia method, and another is the GP method, and the difference between them is expressed as follows. As discussed above, the relationship of the angular size of compact sources with redshifts in Eq\ref{eq2} and CDDR in Eq\ref{eq1} can deprive the luminosity distance of QSO:
\begin{equation}
D_{L,QSO} = \frac{lm(1+z)^2}{\theta_{obs}}
\end{equation}
where $\theta_{obs}$ is the observation angular size of quasars. As we have contact with two samples at the same redshift, we can alternate the luminosity distance of SNe Ia with the quasars' because of the relation in Eq\ref{eq2}. where $\tau(z)$ has the parameter that we need to constraint. After all, what can not be ignored is that the error of the modulus is not only from the observation but also the luminosity distance of quasars. The error of observational angular size and the linear size of the compact structure $l_m$ will transmit to the luminosity then modulus. Considering the error transfer formula, the error of modulus can be written as
\begin{equation}
\sigma_{\mu_{th}}^{2} = (\frac{\partial \mu_{th}}{\partial lm})^{2}lm_{err}^2 + (\frac{\partial \mu_{th}}{\partial \theta})^2\theta_{err}^2
\end{equation}
  where $lm_{err}$ is $0.25\times10^{-6}$, and $\theta_{err}$ is the observation error of angular size. Finally, the total error of modulus is their quadratic sum.
\begin{equation}
\sigma_{total}^2 = \sigma_{\mu_{th}}^{2}+\sigma_{\mu_{obs}}^{2}
\end{equation}
where the $\sigma_{\mu_{th}}$ represents the corresponding error given by quasars, and the $\sigma_{\mu_{obs}}$ represent the error of the distance modulus given by the observational data. The $\sigma_{total}$ represents the total error of modulus distance.

Now the likelihood estimator is determined by $\chi^2$ statistics
\begin{equation}
\chi^2 = \frac{(\mu_{th}-\mu_{obs})^2}{\sigma_{total}^2}
\label{eq4}
\end{equation}
where $\mu_{obs}$ represent the distance modulus given by the observational data, and $\mu_{th}$ represent the distance modulus. In this paper, we adopt the observation quantities ($\theta,\mu,z$) from those Pantheon samples and compact radio quasars sample that we have disposed formerly to constrain the cosmic opacity ($\tau(z)$). By constraining the parameter of $\tau(z)$, one can justify the cosmic opacity cosmology independently and whether the parameter influences the cosmic opacity or not. By evaluating the $\chi^2$ function defined in Eq\ref{eq4} with the Markov Chain Monte Carlo (MCMC) method, we can obtain the best-fit values and the errors of each model parameter. The MCMC method that has been widely used in cosmological studies makes it easy to study the effects of systematic uncertainties comprehensively with a simple inclusion of priors. We use the public python module of emcee \citep{Foreman2013}.

\subsection{Constraints with the nearby SNe Ia method }
\begin{figure}[h]
  \begin{minipage}[t]{0.48\linewidth}
  \centering
   \includegraphics[width=40mm]{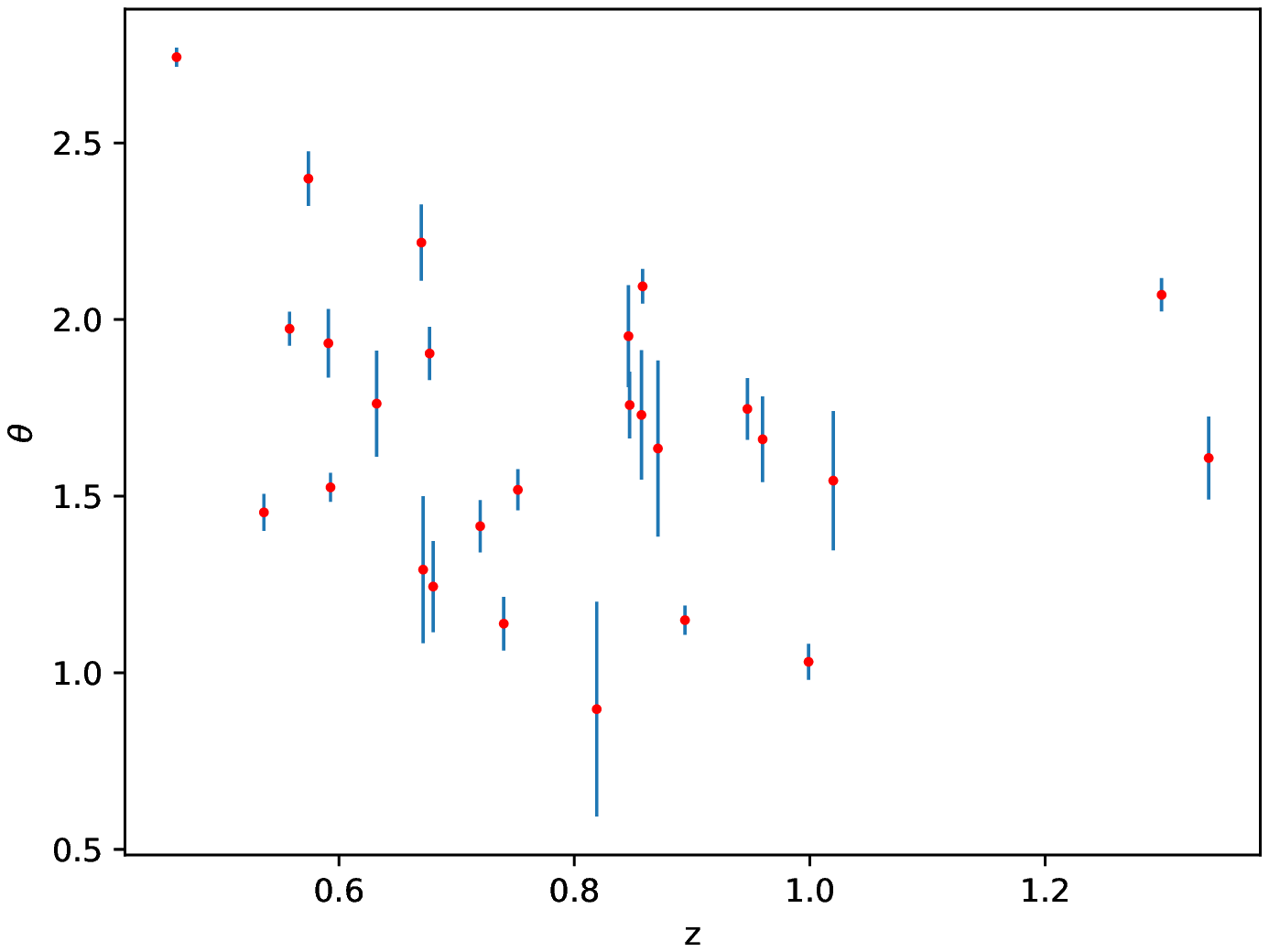}
   \includegraphics[width=40mm]{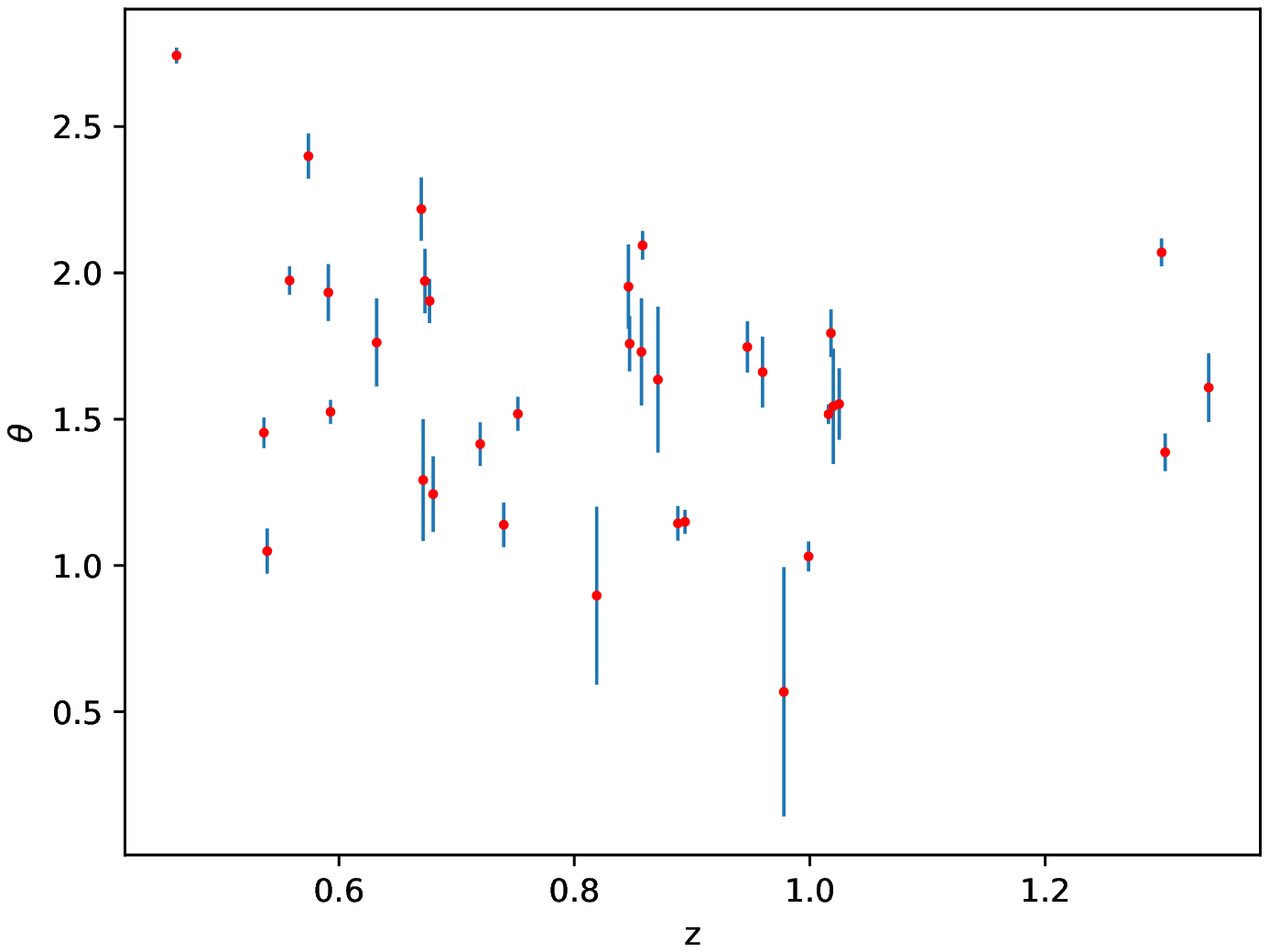}
   \end{minipage}
  \begin{minipage}[t]{0.48\linewidth}
  \centering
   \includegraphics[width=40mm]{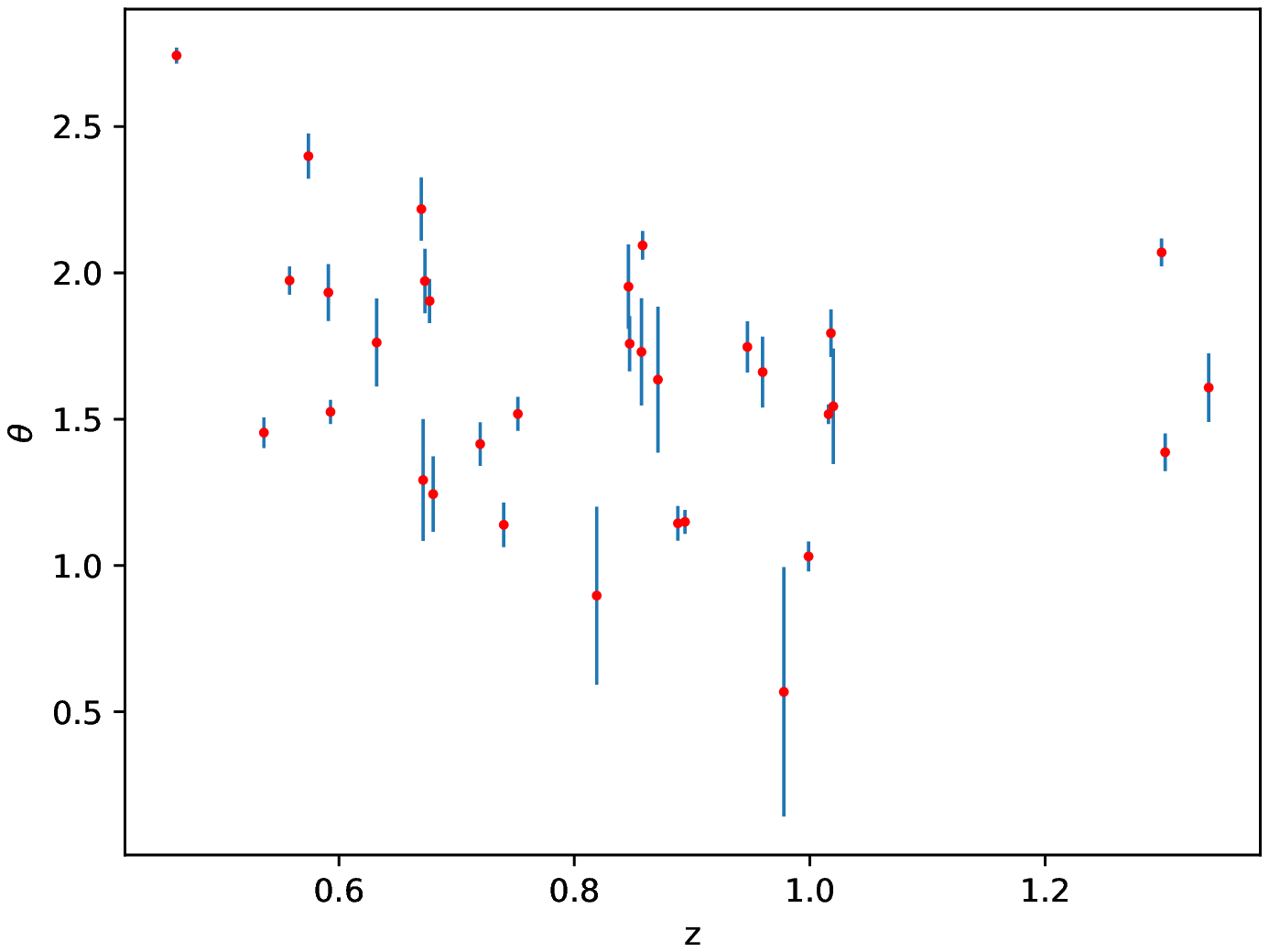}
   \includegraphics[width=40mm]{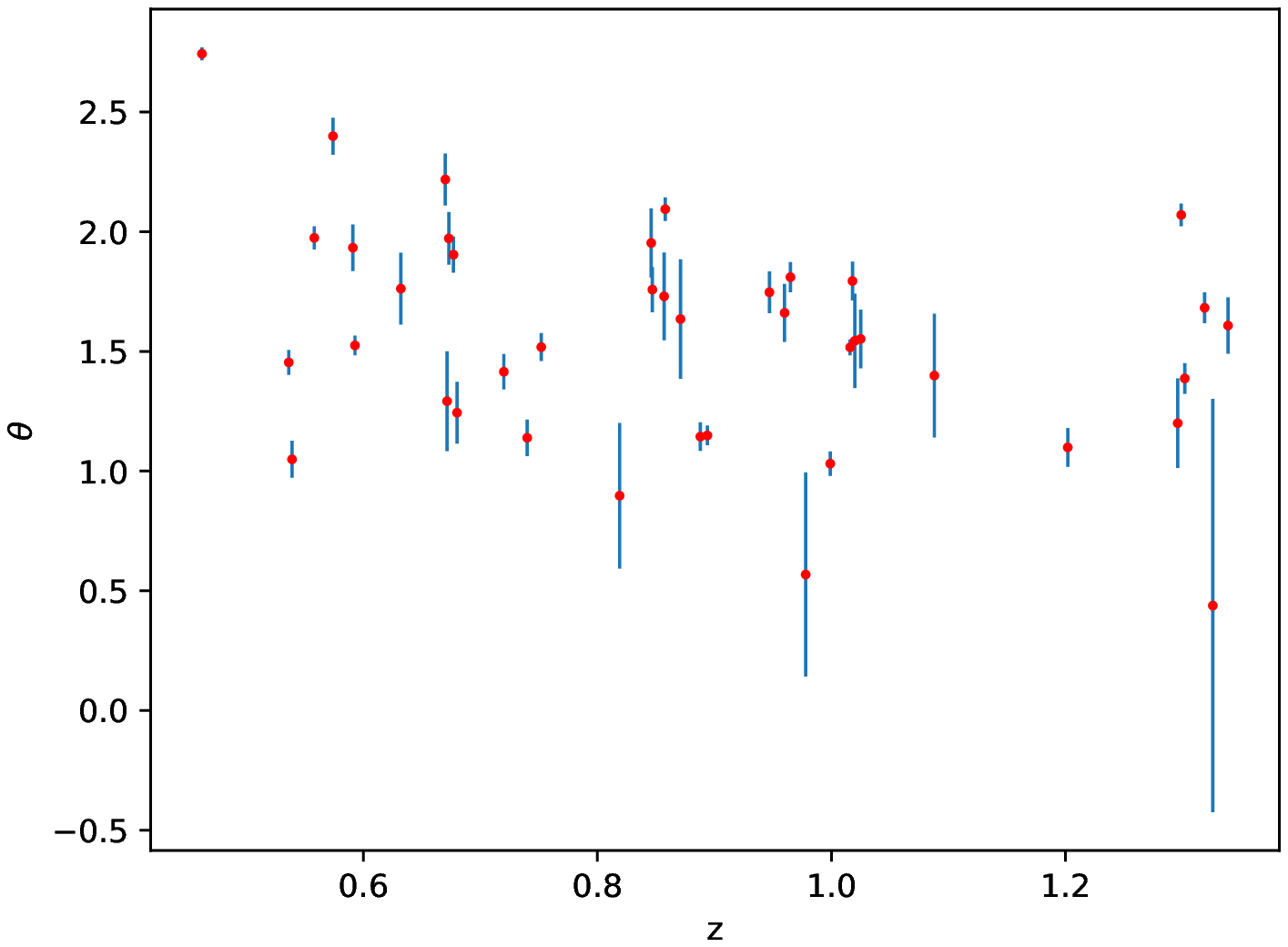}
  \end{minipage}%
    \caption{\label{fig1}{The four groups data that are obtained through the nearby SNe Ia method.} }
\end{figure}

The nearby SNe Ia method is the general way using a redshift matching criterion $\Delta z < d (d \in [0.002,0.005])$ to match two samples at the same redshift. It was verified that it is unrealistic to decrease $\Delta z$ below the total $1\sigma$ error of observational redshifts $\sigma_{total}=\sigma_{z,SN}+\sigma_{z,QSO}= 0.002$ because of the limited accuracy of the redshifts of observations \citep{Qi2019}. Therefore in this paper, we choose four criteria that $d$ is supplied from 0.002 to 0.005. We get four groups of data that matched quasars data with 120 samples and pantheon samples with 1048 SN Ia. The size of the four groups are shown in Figure \ref{fig1}. We can see the bigger matching criterion the more samples from matched data. When $\Delta z < 0.005$, there is the largest group with 41 samples, and the more samples give the better accuracy.

\begin{figure}[t]
\centering
{
\includegraphics[width=8cm]{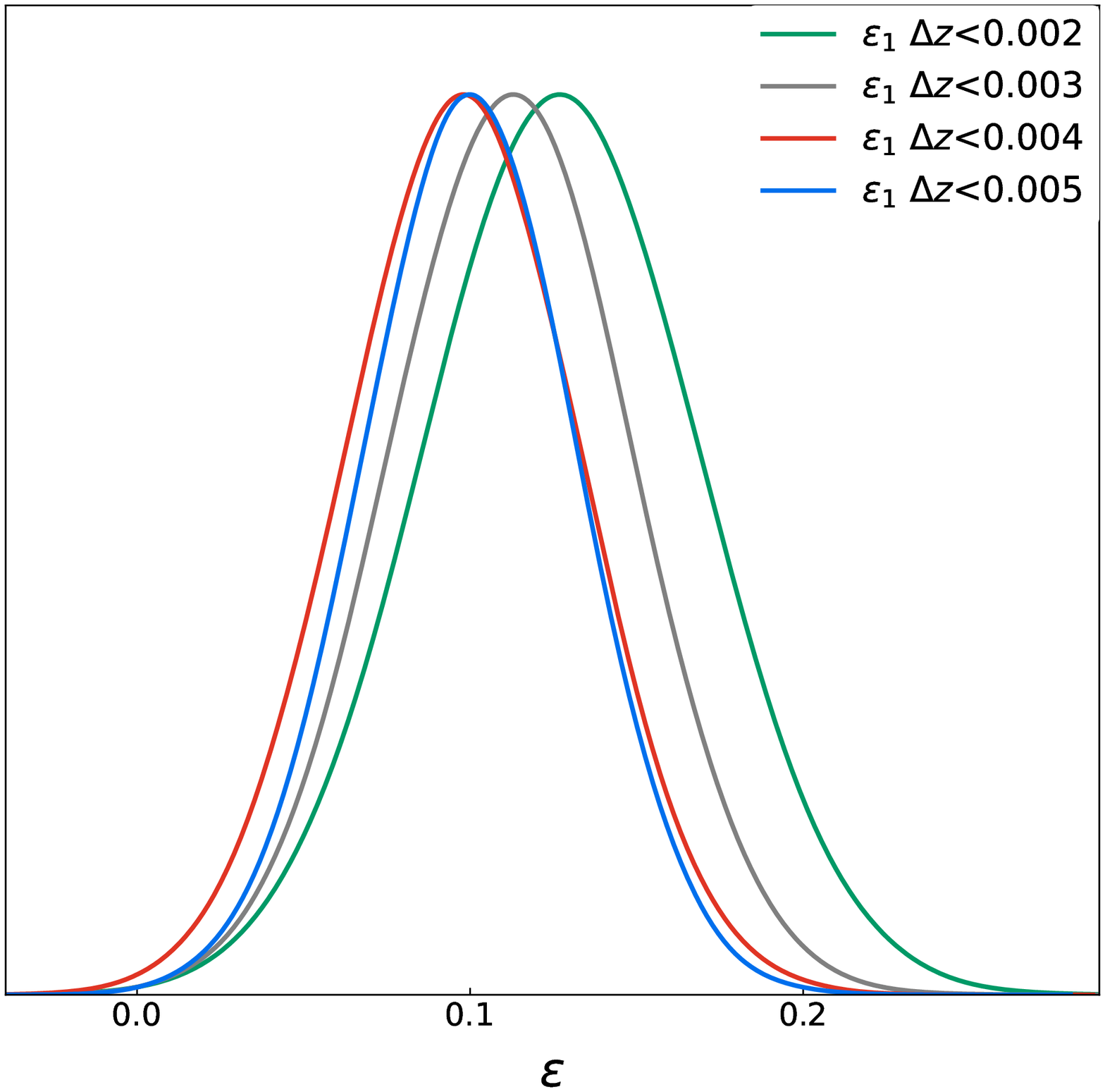}
}
\caption{The distribution of the $F_1$ cosmic opacity parameter $\epsilon$ deprived from the nearby SNe Ia method.}
\label{Figure1}
\end{figure}

\begin{figure}[htbp]
\centering
{
\includegraphics[width=8cm]{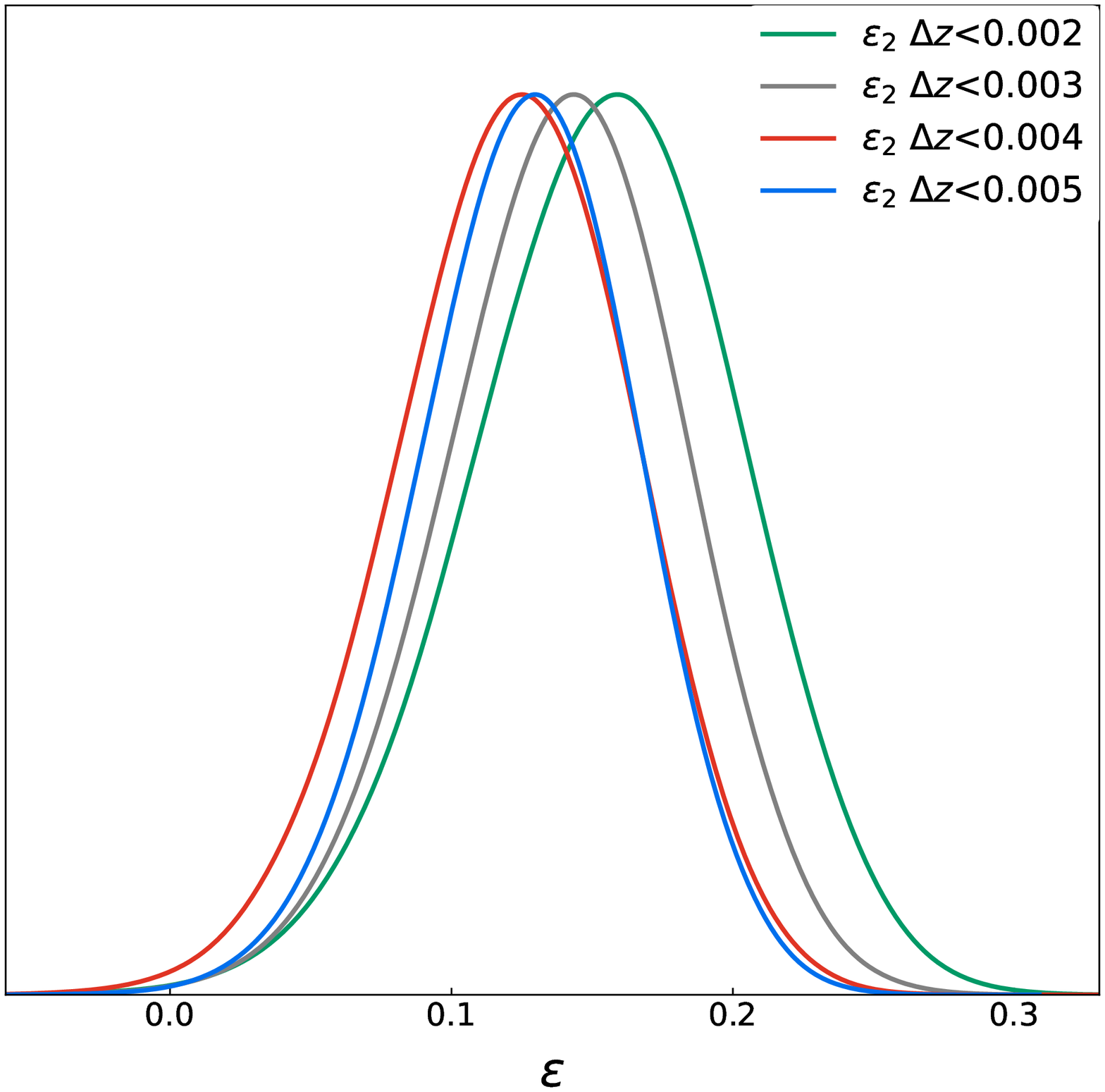}
}
\caption{The distribution of the $F_2$ cosmic opacity parameter $\epsilon$ deprived from the nearby SNe Ia method.}
\label{Figure2}
\end{figure}

Based on those discussions above, we have gotten four groups of data so that we have four results for each parameterized function. The best-fitted parameter values, and corresponding $1\sigma$ standard deviation is listed in Table \ref{Table1} and their distribution are shown in Figures \ref{Figure1} and \ref{Figure2}. As shown in Table \ref{Table1}, the best-fit $\epsilon$ parameter of two functions through the first method are both best constrained under the matching criterion $\Delta z < 0.005$: for the $F_1$ function, the best-fit $\epsilon$ parameter with $1\sigma$ confidence level is $\epsilon_1=0.100\pm0.032$, and for $F_2$ function it is $\epsilon_2=0.126^{+0.041}_{-0.037}$. Comparing the results of a function under different matching criteria, one can see that it is essential to choose an appropriate $\Delta z$. Focus on the result of the $F_1$ function seems to become more accurate as of the matching criterion is upper to 0.005, which has the minimum standard deviation. This trend applies to the $F_2$ function as well. Therefore the quantity of data affects the constraints on parameters significantly and the larger data will improve the accuracy of the constraint on model parameters. It is important to concern the difference between the two assumed $\tau(z)$ parametrizations. The results deprived under different $\Delta z$ of two parametrization functions are shown in Table \ref{Table1}. Comparing the constraints on the two functions under the same $\Delta z$, we can see that the choice of parametrization impacts the results.

\begin{table}
\begin{center}
\caption[]{ The best-fit values and their $1\sigma$ standard errors of the cosmic opacity  parameter $\epsilon$ constrained by the Pantheon and quasars data through the nearby SNe Ia and GP methods.}\label{Table1}

%%Please Capitalize the First Letter of Each Notional Word in table's caption

 \begin{tabular}{clcl}
  \hline\noalign{\smallskip}
  & $F_1$   & $F_2$       \\
  \hline\noalign{\smallskip}
$\Delta z<0.002$    & 0.127$\pm0.042$    & $0.154^{+0.050}_{-0.045}$ \\
$\Delta z<0.003$    & 0.112$\pm0.036$    & $0.140^{+0.045}_{-0.040}$ \\
$\Delta z<0.004$    & 0.098$\pm0.036$Csaki2002    & $0.122^{+0.045}_{-0.040}$ \\
$\Delta z<0.005$    & 0.100$\pm0.032$    & $0.126^{+0.041}_{-0.037}$ \\ \hline
    GP              & -0.005$\pm0.013$   & $-0.018\pm0.018$\\
  \noalign{\smallskip}\hline
\end{tabular}
\end{center}
\end{table}

\begin{figure}[ht]
\centering
{
\includegraphics[width=8cm]{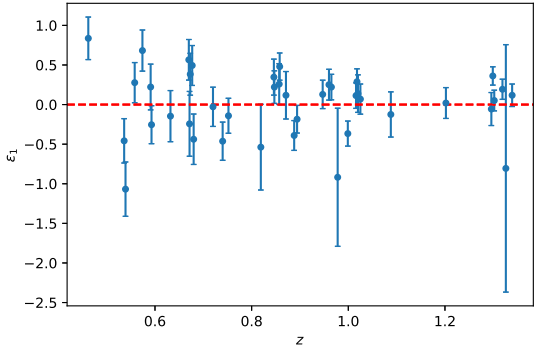}
}
\caption{the $\epsilon_{1}$ and their errors are directly calculated by the formula through the matching data using the nearby SNe Ia way when $\Delta z < 0.005$  in $F_1$ model.}
\label{e1}
\end{figure}

\begin{figure}[ht]
\centering
{
\includegraphics[width=8cm]{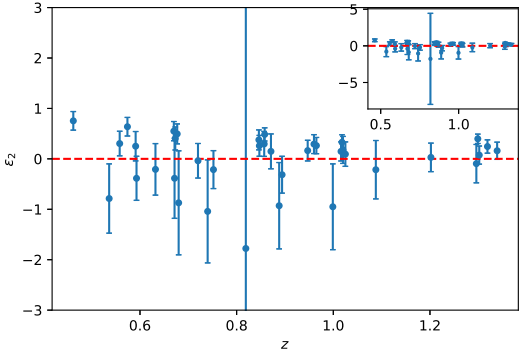}
}
\caption{the $\epsilon_{2}$ and their errors are directly calculated by the formula through the matching data using the nearby SNe Ia way when $\Delta z < 0.005$  in $F_2$ model.}
\label{e2}
\end{figure}

Considering that there is not much data satisfying the matching conditions, we may not be able to accurately represent the distribution of $\epsilon_{1}$ and $\epsilon_2$ by the Monte Carlo method, we bring $\Delta z < 0.005$ of the data directly into the formula calculation, that is, every set of supernova and quasar data that meet the matching conditions can be used to represent an $\epsilon_{1}$ and $\epsilon_2$, the errors of $\epsilon_{1}$ and $\epsilon_2$ are given by error transfer. As a result, we get the error diagrams for $F_1$ and $F_2$, respectively, as shown in Figure \ref{e1} and Figure \ref{e2}. As shown in the figure, in the case of the $F_1$ model, there are 17 data that are consistent with ($\epsilon_{1}=0$) within the error range. In the case of the $F_2$ model, 19 data are consistent with ($\epsilon_{2}=0$) within the error range.

\subsection{Constraints with the GP method}

Another method used the python package GaPP to reconstruct QSO angular size as a function of redshift from the binned data, similar to the method applied in \citep{2021EPJC...81..903L}. The python package used to reconstruct the QSO data was based on Gaussian Processes (GP) \citep{Seikel2012} and depended on the mean function and the covariance function $k(x,x^{~})$. \cite{yan2019exploring,Zheng2020} have discussed the influence of different covariance function and mean function, which demonstrated their insignificant effects on the final reconstruction results. Therefore we choose the zero mean function and the Cauchy covariance function.
\begin{equation}
k(x,\tilde{x}) = \sigma^2_f \frac{l}{(x-\tilde{x})^2+l^2}
\end{equation}
where $l$ and $\sigma_f$ are hyperparameters that control the typical scale in the x-direction and the amplitude of deviation from the mean function, respectively.

\begin{figure}[ht]
\centering
{
\includegraphics[width=8cm]{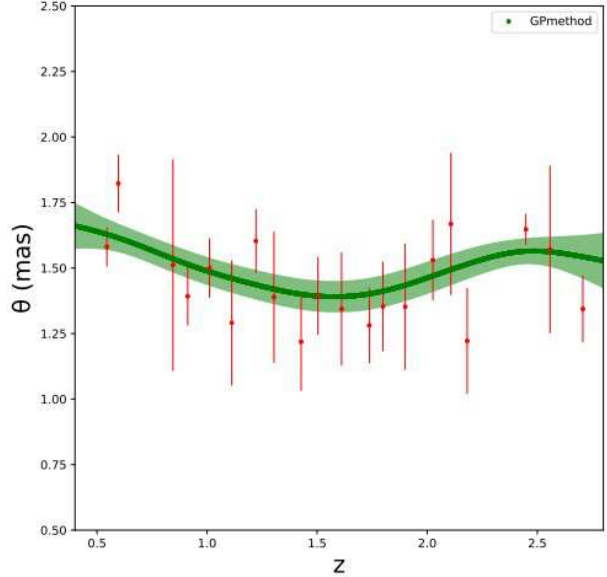}
}
\caption{The reconstructed data of QSO through GP method and the red line reflects the 20 redshift bins.}
\label{fig4}
\end{figure}

\begin{figure}[ht]
\centering
{
\includegraphics[width=6cm]{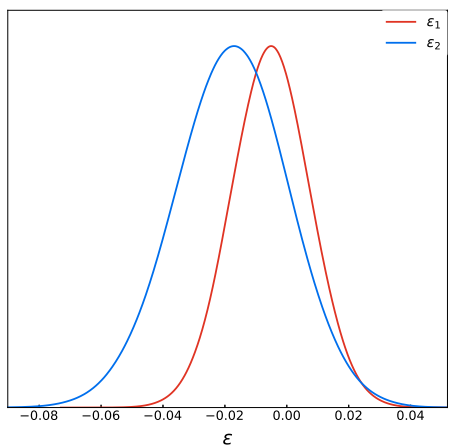}
}
\caption{The distribution of the result tested through GP method.}
\label{Figure3}
\end{figure}

At first, we acquire the binned data from the 120 quasars by grouping them into 20 redshift bins. Then we use the GP method to reconstruct the 20 data to 3000 samples which helps us achieve a one-to-one matching between QSO and SN Ia at the same redshift. And the reconstructed data of QSO is shown in Figure \ref{fig4}. But it turns out difficult to match the whole 1048 SN Ia samples with the reconstructed data because the 120 QSO samples aren't cover low redshift. To match the lowest redshift ($z=0.256$) of QSO, we choose a subset of the pantheon data in which the redshifts are wholly bigger than 0.4. As a result, We obtained 282 SN Ia sub-samples and successfully matched QSO samples one-to-one, and then we constrain the two model function parameters ($\epsilon_1, \epsilon_2$) from those sub samples through the MCMC algorithm.
The results are shown in Figure \ref{Figure3} and Table \ref{Table1}. The best-fit cosmic-opacity parameter with $1\sigma$ confidence level for $F_1$ and $F_2$ functions are $\epsilon_1=-0.005\pm0.013$ and $\epsilon_2=-0.018\pm0.018$ respectively.
We can see that each $1\sigma$ errors are much small than those from the nearby SNe Ia method.
From the overall results, it is also worth comparing the results of the two cosmological-model-independent methods, which shows that the larger the data set, the higher the precision of the parameters. Besides, the advantage of using the GP method is that the data will cover a wider redshift (0.4-2.26). It not only increases the quantity but also provides a higher redshift, which motivates us to study the CDDR in the early universe.

\begin{figure}[ht]
\centering
{
\includegraphics[width=8cm]{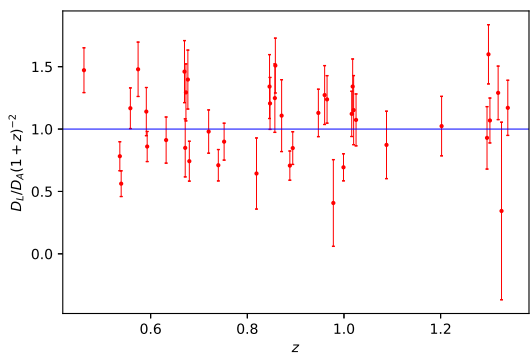}
}
\caption{Direct test of CDDR by nearby method ($\Delta z<0.005$).}
\label{nearbyeta}
\end{figure}

\begin{figure}[ht]
\centering
{
\includegraphics[width=8cm]{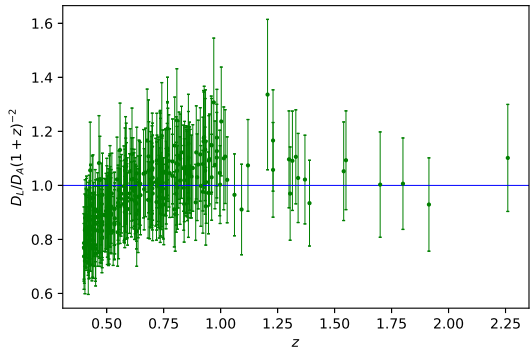}
}
\caption{Direct test of CDDR by GP method.}
\label{GPeta}
\end{figure}

After discussing our results, it is worth comparing them with some test that has been already studied with the different astrophysical probe previously. Constraint results for parameterized formal parameters of CDDR are listed in Table \ref{Table2}. Qi et al. \cite{Qi2019} combined the two SNe Ia data (Joint Light-curve Analysis (JAL) and Pantheon sample) and the simulated gravitational waves from the third-generation gravitational wave detector (the Einstein Telescope, ET). The results show no significant evidence that CDDR is incompatible with the real universe at much higher redshifts ($z\sim2.26$). Their results show that the accuracy of the parameters in the parameterized form of CDDR in 1$\sigma$ confidence is $\Delta\epsilon_1=0.018$ and $\Delta\epsilon_2=0.027$. In addition, they still find that the different parametrizations slightly affects the test of CDDR. Another test done by \cite{Liao2013} combined the 28 observation Hubble parameter data ($H(z)$) and the SNe Ia samples (Union2.1) \citep{Suzuki2012}. They proposed three model-independent methods, including the nearby SNe Ia method that similar to our first method, the interpolation method, and the smoothing method. Their results show a favor to the CDDR ($\Delta\epsilon_1=0.10$ and $\Delta\epsilon_2=0.12$). They also find that the results are not sensitive to the method. And those methodologies was applied and extended in many research \citep{Liao2015}. Li et al. \cite{Li2013} constrained the parameter by providing $D_L$ from the Union2.1 type Ia supernova sample and $D_A$ from two cluster samples \citep{Bonamente2006,Sereno2006}. By comparing our result of the nearby SNe Ia method with those from \cite{Li2013}, we can see that we get smaller error bars when using larger SNe Ia data (Pantheon samples). And in contrast to the similar negative best-fit value constrained by \cite{Liao2013}, our result given through the GP method obtain a smaller error bar too. As to the other three methods, the difference between them is weak, but the GP method shows a huge progression. In addition, we tested CDDR directly using data from the nearby and GP methods, respectively, as shown in Fig.\ref{nearbyeta} and Fig.\ref{GPeta}. Therefore our results foresee the population that the CDDR is constrained by these cosmological-model-independent ways specially GP method.

\begin{table}
\begin{center}
\caption[]{ The best values and their 1 $\sigma$ errors deprived from different observational data and method.}\label{Table2}

%%Please Capitalize the First Letter of Each Notional Word in table's caption

 \begin{tabular}{clcl}
  \hline\noalign{\smallskip}
&$F_1$ &$F_2$\\
  \hline\noalign{\smallskip}
$\Delta z<0.005$      Pantheon+Quasars  & $0.100^{+0.032}_{-0.032}$  & $0.126^{+0.041}_{-0.037}$ \\
GP            Pantheon+Quasars  & -0.005$^{+0.013}_{-0.013}$ & $-0.018^{+0.018}_{-0.018}$ \\ \hline
$\Delta z<0.005$      Pantheon+GW (ET) \citep{Qi2019}     & $0.009^{+0.018}_{-0.018}$ & $0.013^{+0.027}_{-0.027}$\\
                      Union2.1+Cluster \citep{Li2013}  & $0.009^{+0.059}_{-0.055}$ & $0.014^{+0.071}_{-0.069}$\\ \hline
interpolation Union2.1+H(z) \citep{Liao2013}   & $-0.01^{+0.10}_{-0.10}$ & $-0.01^{+0.12}_{-0.12}$ \\
  \noalign{\smallskip}\hline
\end{tabular}
\end{center}
\end{table}

\section{Conclusions}

\label{sect:conclusion}
The Cosmic distance duality relation plays an important role in the universe. It connects the luminosity distance and the angular diameter distance simply, which is a basic relationship. The effectiveness of CDDR has a direct impact on observations, such as photon number conservation and gravity correction theory. Therefore, as a basic and important relationship, it is necessary to test CDDR.
In this paper, we mainly constrain two parameterized forms of CDDR ($\epsilon_1$, $\epsilon_2$) from the Pantheon and quasars data through two cosmological-model-independent ways (the nearby SNe Ia and GP methods), and make a comparative analysis.

The release of the new largest SNe Ia data (Pantheon data) with 1048 samples improves the precision of the study of the CDDR, and another advantage that is observed at many high redshifts help us to realize the CDDR in the early universe. We also apply the compact radio quasars sample with 120 samples to replace model-dependent distance in the nearby SNe Ia and GP methods. Compared with the previous work, our innovation is to use the observed large sample data of quasars in the real electromagnetic band with a higher redshift.

In this paper, we make a combination with compact radio quasars data (QSO) and the new largest type Ia supernova (SNe Ia) to test the CDDR through two cosmological-model-independent ways. The first method is the nearby SNe Ia method which use the redshift criterion to match the QSO data and SNe Ia data, and we get four groups of data for different $\Delta z$. The second method is the GP method, and to have a one-to-one match, we use a sub of the SNe Ia data. The two redshift-dependent parametrization function:$\tau(z) = 2\epsilon_1 z$ and $\tau(z) = (1+z)^{2\epsilon_2}-1$ are proposed to express the CDDR. The purpose of our research is to what extent one can test the CDDR if we use two different methods. Finally, we summarize our main conclusion as follows.

For qualitative analysis, when we analyze the observational data with the GP method for $F_1$ and $F_2$, the result of parameters constrained is 0.013 and 0.018, respectively. Although when we use the nearby SNe Ia method to test the CDDR, precision test estimates of the parameters have been given, to independently test the results of the CDDR at different redshifts, we also calculate the parameters at individual redshifts. In our results, we present the two parameters $\epsilon_1$ and $\epsilon_2$ in the form of a scatter graph. Our results show that the CDDR is slightly sensitive to the assumed form of parameterization. It is essential to choose the appropriate parameterized format for $\tau(z)$ when testing the CDDR. As the number of matching criteria adds up, the accuracy of the best-fit parameter value is improved obviously. The GP method shows a prominent improvement in test accuracy. When we concerned about the number of the data, we can find that the larger data will improve the accuracy of the constraint on model parameters. Using the GP method, we can match as much data as possible between supernovae and quasars. Therefore we attain the largest matching data and the most accurate parameter value. As a consequence, the importance of the GP method should be fully taken into account in future investigations.

\begin{acknowledgements}

This work is supported by the National Natural Science Foundation of China under grant Nos. 12105032, 12021003, 11690023, and 11920101003;
the Strategic Priority Research Program of the Chinese Academy of Sciences, Grant No. XDB23000000; Chongqing Science and technology research project No. KJ111206; Natural Science Foundation of Chongqing (Grant No. cstc2021jcyj-msxmX0481);
the Scientific Research and Innovation Project of Graduate Students in Chongqing Nos. CYS20272, CYS21327; and the Interdiscipline Research Funds of Beijing Normal University. We also gratefully acknowledge the financial support from
Funda\c{c}\~ao de Amparo \`a Pesquisa do Estado de S\~ao Paulo (FAPESP),
Funda\c{c}\~ao de Amparo \`a Pesquisa do Estado do Rio de Janeiro (FAPERJ),
Conselho Nacional de Desenvolvimento Cient\'{\i}fico e Tecnol\'ogico (CNPq),
Coordena\c{c}\~ao de Aperfei\c{c}oamento de Pessoal de N\'ivel Superior (CAPES).

\end{acknowledgements}
%\begin{acknowledgements}
%\end{acknowledgements}


%\bibliographystyle{elsarticle-harv}
\bibliographystyle{spphys}

\bibliography{CJP}

\begin{thebibliography}{10}
\providecommand{\url}[1]{{#1}}
\providecommand{\urlprefix}{URL }
\expandafter\ifx\csname urlstyle\endcsname\relax
  \providecommand{\doi}[1]{DOI \discretionary{}{}{}#1}\else
  \providecommand{\doi}{DOI \discretionary{}{}{}\begingroup
  \urlstyle{rm}\Url}\fi

\bibitem{Etherington1933}
I.~Etherington, The London, Edinburgh, and Dublin Philosophical Magazine and
  Journal of Science \textbf{15}(100), 761 (1933).
\newblock \doi{10.1080/14786443309462220}

\bibitem{cao2011the}
S.~Cao, Z.~Zhu, Science China-Physics Mechanics \& Astronomy \textbf{54}(12),
  2260 (2011).
\newblock \doi{10.1007/s11433-011-4559-7}

\bibitem{cao2016testing}
S.~Cao, M.~Biesiada, X.~Zheng, Z.~Zhu, Monthly Notices of the Royal
  Astronomical Society \textbf{457}(1), 281 (2016).
\newblock \doi{10.1093/mnras/stv2999}

\bibitem{Holanda2011}
R.~Holanda, J.~Lima, M.~Ribeiro, Astronomy \& Astrophysics \textbf{528}, L14
  (2011).
\newblock \doi{10.1051/0004-6361/201015547}

\bibitem{cao2011testing}
S.~Cao, N.~Liang, Research in Astronomy and Astrophysics \textbf{11}(10), 1199
  (2011).
\newblock \doi{10.1088/1674-4527/11/10/008}

\bibitem{Li2011}
Z.~Li, P.~Wu, H.~Yu, Astrophysical Journal Letters \textbf{729}(1), L14 (2011).
\newblock \doi{10.1088/2041-8205/729/1/L14}

\bibitem{Yang2019}
T.~Yang, R.~Holand, B.~Hu, Astroparticle Physics \textbf{108}, 57 (2019).
\newblock \doi{10.1016/j.astropartphys.2019.01.005}

\bibitem{Fu2019}
X.~Fu, L.~Zhou, J.~Chen, Physical Review D \textbf{99}(8), 083523 (2019).
\newblock \doi{10.1103/PhysRevD.99.083523}

\bibitem{Lima2011}
J.~Lima, J.~Cunha, V.~Zanchin, Astrophysical Journal Letters \textbf{742}(2),
  L26 (2011).
\newblock \doi{10.1088/2041-8205/742/2/L26}

\bibitem{Fu2020}
X.~Fu, J.~Yang, Z.e.a. Chen, European Physical Journal C \textbf{80}(9), 893
  (2020).
\newblock \doi{10.1140/epjc/s10052-020-08479-6}

\bibitem{Fu2021}
X.~Fu, L.~Zhou, J.e.a. Yang, Chinese Physics C \textbf{45}(6), 065104 (2021).
\newblock \doi{10.1088/1674-1137/abf48a}

\bibitem{Zhou2019}
L.~Zhou, X.~Fu, Z.~Peng, J.~Chen, Physical Review D \textbf{100}(12), 123539
  (2019).
\newblock \doi{10.1103/PhysRevD.100.123539}

\bibitem{Liao2016}
K.~Liao, Z.~Li, S.e.a. Cao, Astrophysical Journal \textbf{822}(2), 74 (2016).
\newblock \doi{10.3847/0004-637x/822/2/74}

\bibitem{Lv2016}
M.~Lv, J.~Xia, Physics of the Dark Universe \textbf{13}, 139 (2016).
\newblock \doi{10.1016/j.dark.2016.06.003}

\bibitem{Fu2017}
X.~Fu, P.~Li, International Journal of Modern Physics D \textbf{26}(9), 1750097
  (2017).
\newblock \doi{10.1142/S0218271817500973}

\bibitem{Ruan2017}
A.~Rana, D.~Jain, S.e.a. Mahajan, Journal of Cosmology and Astroparticle
  Physics (7), 010 (2017).
\newblock \doi{10.1088/1475-7516/2017/07/010}

\bibitem{Li2018}
X.~Li, H.~Lin, Monthly Notices of the Royal Astronomical Society
  \textbf{474}(1), 313 (2018).
\newblock \doi{10.1093/mnras/stx2810}

\bibitem{Lin2018}
H.~Lin, M.~Li, X.~Li, Monthly Notices of the Royal Astronomical Society
  \textbf{480}(3), 3117 (2018).
\newblock \doi{10.1093/mnras/sty2062}

\bibitem{Ruan2018}
C.~Ruan, F.~Melia, T.~Zhang, Astrophysical Journal \textbf{866}(1), 31 (2018).
\newblock \doi{10.3847/1538-4357/aaddfd}

\bibitem{Qi2019b}
J.~Qi, S.~Cao, C.e.a. Zheng, Physical Review D \textbf{99}(6), 063507 (2019).
\newblock \doi{10.1103/PhysRevD.99.063507}

\bibitem{Liao2013}
K.~Liao, Z.~Li, J.~Ming, Z.~Zhu, Physics Letters B \textbf{718}(4-5), 1166
  (2013).
\newblock \doi{10.1016/j.physletb.2012.12.022}

\bibitem{Holanda2013}
R.~Holanda, J.~Carvalho, J.~Alcaniz, Journal of Cosmology and Astroparticle
  Physics (4), 027 (2013).
\newblock \doi{10.1088/1475-7516/2013/04/027}

\bibitem{Liao2015}
K.~Liao, A.~Avgoustidis, Z.~Li, Physical Review D \textbf{92}(12), 123539
  (2015)

\bibitem{Zheng2020}
X.~Zheng, K.~Liao, M.e.a. Biesiada, The Astrophysical Journal \textbf{892}(2),
  103 (2020).
\newblock \doi{10.3847/1538-4357/ab7995}

\bibitem{geng2020gravitational-wave}
S.~Geng, S.~Cao, T.e.a. Liu, Astrophysical Journal \textbf{905}(1), 54 (2020).
\newblock \doi{10.3847/1538-4357/abc076}

\bibitem{2021EPJC...81..903L}
T.~Liu, S.~Cao, S.e.a. Zhang, European Physical Journal C \textbf{81}(10), 903
  (2021).
\newblock \doi{10.1140/epjc/s10052-021-09713-5}

\bibitem{Risaliti2019}
G.~Risaliti, E.~Lusso, Nature Astronomy \textbf{3}(3), 272 (2019).
\newblock \doi{10.1038/s41550-018-0657-z}

\bibitem{Cao2017}
S.~Cao, X.~Zheng, M.e.a. Biesiada, Astronomy \& Astrophysics \textbf{606}, A15
  (2017).
\newblock \doi{10.1051/0004-6361/201730551}

\bibitem{Cao2018b}
S.~{Cao}, M.~{Biesiada}, J.~{Jackson}, X.~{Zheng}, Y.~{Zhao}, Z.H. {Zhu},
  Journal of Cosmology and Astroparticle Physics \textbf{2017}(2), 012 (2017).
\newblock \doi{10.1088/1475-7516/2017/02/012}

\bibitem{Cao2018a}
S.~Cao, M.~Biesiada, J.e.a. Qi, European Physical Journal C \textbf{78}(9), 749
  (2018).
\newblock \doi{10.1140/epjc/s10052-018-6197-y}

\bibitem{Kellermann1993}
K.~Kellermann, Nature \textbf{361}(6408), 134 (1993).
\newblock \doi{10.1038/361134a0}

\bibitem{Gurvits1994}
L.~Gurvits, Astrophysical Journal \textbf{425}(2), 442 (1994).
\newblock \doi{10.1086/173999}

\bibitem{MORABITO1985}
D.~Morabito, R.~Preston, J.e.a. Williams,  \textbf{90}(9), 229–237 (1985).
\newblock \doi{10.1086/113869}

\bibitem{Naess2014}
S.~Naess, M.~Hasselfield, J.e.a. McMahon, Journal of Cosmology and
  Astroparticle Physics (10), 007 (2014).
\newblock \doi{10.1088/1475-7516/2014/10/007}

\bibitem{JACKSON2006}
J.~Jackson, A.~Jannetta, Journal of Cosmology and Astroparticle Physics (11),
  002 (2006).
\newblock \doi{10.1088/1475-7516/2006/11/002}

\bibitem{2015ApJ...806...66C}
S.~{Cao}, M.~{Biesiada}, X.~{Zheng}, Z.H. {Zhu}, Astrophysical Journal
  \textbf{806}(1), 66 (2015).
\newblock \doi{10.1088/0004-637X/806/1/66}

\bibitem{zheng2017ultra}
X.~Zheng, M.~Biesiada, S.e.a. Cao, Journal of Cosmology and Astroparticle
  Physics (10), 030 (2017).
\newblock \doi{10.1088/1475-7516/2017/10/030}

\bibitem{Qi2017}
J.~Qi, S.~Cao, M.e.a. Biesiada, European Physical Journal C \textbf{77}(8), 502
  (2017).
\newblock \doi{10.1140/epjc/s10052-017-5069-1}

\bibitem{Xu2017}
T.~Xu, S.~Cao, J.e.a. Qi, Journal of Cosmology and Astroparticle Physics (6),
  042 (2018).
\newblock \doi{10.1088/1475-7516/2018/06/042}

\bibitem{2020ApJ...888L..25C}
S.~Cao, J.~Qi, M.e.a. Biesiada, Astrophysical Journal Letters \textbf{888}(2),
  L25 (2020).
\newblock \doi{10.3847/2041-8213/ab63d6}

\bibitem{cao2019milliarcsecond}
S.~Cao, J.~Qi, M.e.a. Biesiada, Physics of the Dark Universe \textbf{24},
  100274 (2019).
\newblock \doi{10.1016/j.dark.2019.100274}

\bibitem{2021MNRAS.503.2179Q}
J.~Qi, J.~Zhao, S.e.a. Cao, Monthly Notices of the Royal Astronomical Society
  \textbf{503}(2), 2179 (2021).
\newblock \doi{10.1093/mnras/stab638}

\bibitem{Scolnic2017}
D.~Scolnic, D.~Jones, A.e.a. Rest, Astrophysical Journal  (2017).
\newblock \doi{10.3847/1538-4357/aab9bb}

\bibitem{Qi2019}
J.~Qi, S.~Cao, Y.~Pan, J.~Li, Physics of the Dark Universe \textbf{26}, 100338
  (2019).
\newblock \doi{10.1016/j.dark.2019.100338}

\bibitem{Foreman2013}
D.~Foreman-Mackey, D.~Hogg, D.~Lang, J.~Goodman, Publications of the
  Astronomical Society of the Pacific \textbf{125}(925), 306 (2013).
\newblock \doi{10.1086/670067}

\bibitem{Seikel2012}
M.~Seikel, C.~Clarkson, M.~Smith, Journal of Cosmology and Astroparticle
  Physics (6), 036 (2012).
\newblock \doi{10.1088/1475-7516/2012/06/036}

\bibitem{yan2019exploring}
Y.~Wu, S.~Cao, J.e.a. Zhang, Astrophysical Journal \textbf{888}(2), 113 (2020).
\newblock \doi{10.3847/1538-4357/ab5b94}

\bibitem{Suzuki2012}
N.~Suzuki, D.~Rubin, C.e.a. Lidman, Astrophysical Journal \textbf{746}(1), 85
  (2012).
\newblock \doi{10.1088/0004-637x/746/1/85}

\bibitem{Li2013}
Z.~Li, P.~Wu, H.~Yu, Z.~Zhu, Physical Review D \textbf{87}(10), 103013 (2013).
\newblock \doi{10.1103/PhysRevD.87.103013}

\bibitem{Bonamente2006}
M.~Bonamente, M.~Joy, S.e.a. LaRoque, Astrophysical Journal \textbf{647}(1), 25
  (2006).
\newblock \doi{10.1086/505291}

\bibitem{Sereno2006}
M.~Sereno, E.~Filippis, G.~Longo, M.~Bautz, Astrophysical Journal
  \textbf{645}(1), 170 (2006).
\newblock \doi{10.1086/503198}

\end{thebibliography}

\end{document}